\documentclass[floatfix,reprint,amsmath,amssymb,aps]{revtex4-2}
\usepackage{CJK}
\usepackage{graphicx}
\usepackage{color}
\usepackage{amsmath}
\usepackage{amssymb}
\usepackage{amsthm}
\usepackage{extarrows}
\usepackage{tabularx}
\usepackage{dsfont}
\usepackage{booktabs}
\usepackage{float}
\usepackage{lineno}

\usepackage{appendix}

\usepackage{xcolor}

\usepackage[dvipdfm,colorlinks,urlcolor=blue,linkcolor=blue,anchorcolor=blue,citecolor=blue]{hyperref}

\begin{document}

\title{Gravity-Induced Modulation of Negative Differential Thermal Resistance in Fluids}

\author{Qiyuan Zhang$^1$}
 \author{Juncheng Guo$^1$}
\author{Juchang Zou$^1$}
\author{Rongxiang Luo$^{1,2}$}
\email{phyluorx@fzu.edu.cn}



\affiliation{
$^1$ Department of Physics, Fuzhou University, Fuzhou 350108, Fujian, China\\
$^2$ Fujian Science and Technology Innovation Laboratory for Optoelectronic Information of China, Fuzhou 350108, Fujian, China
}

\date{\today }

\begin{abstract}
We investigate how gravity influences negative differential thermal resistance (NDTR) in fluids modeled by multiparticle collision dynamics. In the integrable case, we derive the heat flux formula for the system exhibiting the NDTR effect, and show that by introducing a gravity along the direction of the thermodynamic force, the temperature difference required for the occurrence of NDTR can be greatly reduced. Meanwhile, we also demonstrate that the heat-bath-induced NDTR mechanism---originally found to be applicable only to weakly interacting systems---can now operate in systems with stronger interactions due to the presence of gravity, and further remains robust even in mixed fluids. These results provide new insights into heat transport and establish a theoretical foundation for designing fluidic thermal devices that harness the NDTR effect under gravity.
\end{abstract}

\maketitle

\textit{Introduction.} The study of negative differential thermal resistance (NDTR)---a phenomenon in which the heat flux decreases with increasing temperature difference between the heat baths---is crucial for uncovering fundamental mechanisms of heat transport and enables novel functionalities in thermal management and energy conversion (see~\cite{Li2012,Landi2022} for reviews and references therein). First predicted in 2004 for nonlinear phononic lattices~\cite{Li2004}, NDTR has since enabled the theoretical design of novel thermal control devices (TCDs)~\cite{Li2006,Wang2007,Wang2008,Wu2012}, including thermal transistors, logic gates, and memory elements. To date, extensive studies based on lattice models have revealed the fundamental characteristics of NDTR, as well as the conditions and mechanisms underlying its emergence~\cite{S2006,Yang2007,He2009,Shao2009,P2010,Zhong2011,M2015}. Furthermore, motivated by these pioneering studies, NDTR has been extensively explored in diverse quantum systems as a route toward quantum TCDs~\cite{Ben2014,Schuab2016,Joulain2016,Ren2013,Fornieri2016,Guo2018,Liu2019,Du2019}, notably the quantum thermal transistor.

Similarly, heat-bath-induced NDTR in fluids has recently attracted significant interest due to its fundamental importance and potential for enabling novel fluidic TCDs~\cite{Luo2019,Luo2022,Zou2025}. In 2019, it was demonstrated that NDTR can be induced in one-dimensional (1D) hard-point gas systems, representing 1D fluids, when coupled to heat baths at different temperatures~\cite{Luo2019}. In such systems, NDTR arises because lowering the temperature of the cold bath suppresses particle motion, thereby drastically reducing the frequency of collisions between the colder particles and the hot bath. Consequently, the heat flux remains small even for large temperature differences across the baths. Subsequently, studies of the mesoscopic fluid model based on multiparticle collision (MPC) dynamics~\cite{Luo2022} further showed that the heat-bath-induced NDTR mechanism also applies to high-dimensional fluidic systems with weak interactions and is very robust even in mixed fluids. These advances in NDTR have established new fundamental insights into a broad class of nonequilibrium fluid systems.\par

Under certain conditions, Earth's gravitational field can significantly influence the thermophysical properties of a broad class of fluid systems~\cite{Thorsten2001,van1997,Scholz2017,van1985,Haussmann1999}, including granular gases, macromolecular gases, and simple fluids near gas-liquid critical points. In such systems, gravity induces an inhomogeneous density distribution, leading to height-dependent variations in all thermophysical properties. A clear understanding of gravitational effects on heat transport in these fluids is therefore essential. The present work addresses a central question: How does gravity modify the emergence and characteristics of NDTR in such systems?\par

To obtain a clear and minimal physical picture about the answer, we consider a two-dimensional (2D) system of particles confined in a rectangular box and governed by MPC dynamics---a stochastic method used to simulate simple fluids. The MPC approach replaces conventional deterministic molecular dynamics with stochastic velocity rotations~\cite{M1999}. A key advantage of this method is its ability to correctly reproduce hydrodynamic behavior with significant computational efficiency~\cite{Gompper2009,Padding2006}. MPC has been applied to study coupled particle and heat transport in nonequilibrium systems~\cite{Benenti2014,Luo2020}, and more recently, has offered fundamental insights into heat conduction mechanisms~\cite{Luo2025}.

Here, we investigate the validity of the heat-bath-induced NDTR mechanism in fluids under gravity. First, we derive an analytical expression for the heat flux that captures the emergence of NDTR in the presence of gravity, and we show that when the gravitational field is aligned with the thermodynamic force, the temperature difference required for the onset of NDTR is significantly reduced. Second, we demonstrate that gravity enables this mechanism to operate in systems with stronger interparticle interactions, thereby extending its applicability beyond the weakly interacting systems for which it was originally established. Finally, we show that the mechanism remains robust in mixed fluids. Additional results for the case in which the gravitational field opposes the thermodynamic force are provided in the Supplemental Material.

\begin{figure}[t]
\centering
\includegraphics[width=3.2cm]{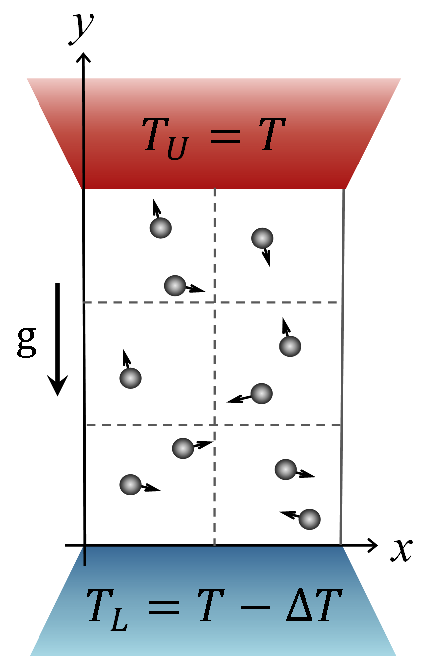}
\caption{Schematic of the 2D fluid model based on multiparticle collision dynamics, consisting of interacting point particles confined in a rectangular domain. Dashed cells denote the spatial grid used for stochastic collision events. A uniform gravitational field $\mathbf{g} = (0, -g)$ is applied along the $y$-axis, and heat baths at fixed temperatures $T_U$ (upper) and $T_L$ (lower) couple to the system at $y = H$ and $y = 0$, respectively (see the main text for further details).}
\label{model}
\end{figure}

\textit{2D fluid model under gravity.} The system, illustrated in Fig.~\ref{model}, comprises $N$ identical point particles of mass $m$, confined to a rectangular domain of length $L$ and height $H$ in the $(x,y)$ plane. The heat baths at fixed temperatures $T_U$ (upper, $y=H$) and $T_L$ (lower, $y=0$) couple to the system at the respective boundaries. Upon reaching a boundary, a particle is reflected with a new velocity $\mathbf{v} = (v_x, v_y)$ drawn from the following distributions~\cite{1978T,T1998}:
\begin{equation}\label{Eqfv}
\begin{aligned}
  f_{\alpha}(v_x) &= \sqrt{\frac{m}{2\pi k_{\mathrm{B}} T_\alpha}} \exp\!\left(-\frac{m v_x^2}{2k_{\mathrm{B}} T_\alpha}\right), \\
  f_{\alpha}(v_y) &= \frac{m |v_y|}{k_{\mathrm{B}} T_\alpha} \exp\!\left(-\frac{m v_y^2}{2k_{\mathrm{B}} T_\alpha}\right),
\end{aligned}
\end{equation}
where $\alpha = U,L$ labels the heat bath, and $k_{\mathrm{B}}$ is Boltzmann’s constant. The sampling enforces the physical constraint that $v_y > 0$ for reflections from $y=0$ and $v_y < 0$ for those from $y=H$; $v_x$ remains unrestricted. Periodic boundary conditions are imposed in the $x$-direction; we note that all results remain unchanged under fixed boundary conditions.

The bulk dynamics are governed by multiparticle collision (MPC) dynamics~\cite{M1999,Gompper2009}. In this approach, time and space are coarse-grained, and deterministic interparticle forces are replaced by stochastic collision rules. The system evolves in discrete time steps of duration $\tau$, each comprising: a \emph{streaming phase}, during which particles move under the uniform gravitational field $\mathbf{g} = (0, -g)$, and an instantaneous \emph{collision phase}, where velocities are updated stochastically within spatial cells. During streaming, the equations of motion are given by Newton's second law, $d\mathbf{v}_i/dt = \mathbf{g}$, with the trivial solution~\cite{Hurtado2020}
\begin{equation}\label{Eq22}
\begin{aligned}
  \mathbf{r}_i(t) &= \mathbf{r}_i(0) + \mathbf{v}_i(0)\,t + \tfrac{1}{2}\mathbf{g}\,t^{2}, \\
  \mathbf{v}_i(t) &= \mathbf{v}_i(0) + \mathbf{g}\,t.
\end{aligned}
\end{equation}
For the collision step, the domain is partitioned into identical square cells of side length $a$ (see Fig.~\ref{model}). Within each cell, the velocity of particle $i$ is updated as
\begin{equation}\label{EqMPC}
  \mathbf{v}_i \;\mapsto\; \mathbf{V}_{\mathrm{c.m.}} + \hat{\mathcal{R}}^{\pm\theta} \bigl( \mathbf{v}_i - \mathbf{V}_{\mathrm{c.m.}} \bigr),
\end{equation}
where $\mathbf{V}_{\mathrm{c.m.}}$ is the cell’s center-of-mass velocity, and $\hat{\mathcal{R}}^{\pm\theta}$ denotes a planar rotation by angle $+\theta$ or $-\theta$, each selected with probability $1/2$. This collision rule ensures the conservation of total momentum and kinetic energy within each cell. A rotation angle $\theta = \pi/2$ maximizes momentum mixing and is commonly adopted for optimal thermalization~\cite{Benenti2014}. Since the collision frequency scales as $\tau^{-1}$, the time step $\tau$ serves as a tunable parameter that controls the effective interaction strength and, consequently, the transport properties (e.g., thermal conductivity~\cite{Luo2025}) of the fluid.

In our model, the heat baths are set to temperatures $T_U = T$ and $T_L = T - \Delta T$, respectively, where $\Delta T$ denotes the applied temperature bias. We work in dimensionless (natural) units~\cite{ga}, with $k_{\mathrm{B}} = T = m = a = 1$, $\theta = \pi/2$, $L = 2$, $H = 10$, and an average particle number density $\rho = N/(LH) = 6$. Initial conditions are prepared by uniformly distributing all particles in the domain and assigning each particle a random velocity drawn from the Maxwell--Boltzmann distribution at temperature $T$. After the system reaches a steady state, the heat flux $J$ is computed as the energy transfer per unit area and unit time between particles and the heat bath. Spatial profiles of the number density $\rho(h)$ and local temperature $T(h)$---where $h$ is the coordinate along the $y$-direction---are measured following the protocol of Refs.~\cite{Luo2019,2020Luo}: unlike the previous case of sampling at unit time intervals, $T(h)$ must now be rigorously computed via the time average
average $T(h)=\lim\limits_{\mathcal{T} \to \infty}\frac{1}{\mathcal{T}}\int_{0}^\mathcal{T}\frac{m \mathbf{v}_{h}^2(t)}{k_{\mathrm{B}}}dt $, since the velocity $\mathbf{v}_{h}(t)$ of a particle at height $h$, as shown in Eq.~(\ref{Eq22}), is time-dependent under gravity. To preserve Galilean invariance in the stochastic collision step, the collision grid is randomly shifted before each collision event~\cite{Ihle2001}. All data points shown in this work represent numerical simulation results. Statistical uncertainties are below $1\%$; consequently, error bars are omitted for clarity.\par


\textit{Theoretical analysis.}
We analytically examine how gravity modifies the properties of NDTR in fluids and elucidate its underlying mechanism. In the noninteracting case---realized when the MPC time step $\tau = \infty$, such that particles traverse the system under gravity alone---the model becomes integrable. For this case, the steady-state heat flux $J$ across the system is given by (see Sec.~I of the Supplemental Material for the derivation~\cite{SM}):
\begin{equation}\label{J11}
  J = \rho H \, \frac{1+d}{2} \left| \frac{k_{\mathrm{B}} (T_U - T_L)}{\langle t_{U\to L} \rangle + \langle t_{L\to U} \rangle} \right|,
\end{equation}
where $d = 1,2,3$ denotes the spatial dimensionality (the result applies equally to one-, two-, and three-dimensional systems). Here, $\langle t_{U\to L} \rangle$ and $\langle t_{L\to U} \rangle$ are the mean transit times for a particle moving from the upper to the lower bath and vice versa, respectively. Explicitly,
\begin{widetext}
\begin{equation}\label{Eqt}
\begin{aligned}
  \langle t_{U\rightarrow L}\rangle & = \frac{1}{g}\sqrt{\frac{\pi k_{\mathrm{B}} T_U}{2m}}\left\{e^{\frac{mgH}{k_{\mathrm{B}}T_U}}\left[1-\mathrm{Erf}(\sqrt{\frac{mgH}{k_{\mathrm{B}}T_U}})\right]-1\right\}+\sqrt{\frac{2H}{g}},\\
  \langle t_{L\rightarrow U}\rangle & =\frac{1}{g}\sqrt{\frac{\pi k_{\mathrm{B}} T_L}{2m}}\left\{e^{\frac{mgH}{k_{\mathrm{B}}T_L}}\left[1+\mathrm{Erf}(\sqrt{\frac{mgH}{k_{\mathrm{B}}T_L}})\right]-1\right\}-\sqrt{\frac{2H}{g}}.
\end{aligned}
\end{equation}
\end{widetext}
In the limit of $g \to 0$, a Taylor expansion of Eq.~(\ref{Eqt}) yields the gravity-free results of Ref.~\cite{Luo2019}: $\langle t_{U\rightarrow L}\rangle= H \sqrt{\pi m/2k_{\mathrm{B}}T_U}$ and $\langle t_{L\rightarrow U}\rangle= H \sqrt{\pi m/2k_{\mathrm{B}}T_L}$.

\begin{figure}
\centering
\includegraphics[width=7.5cm]{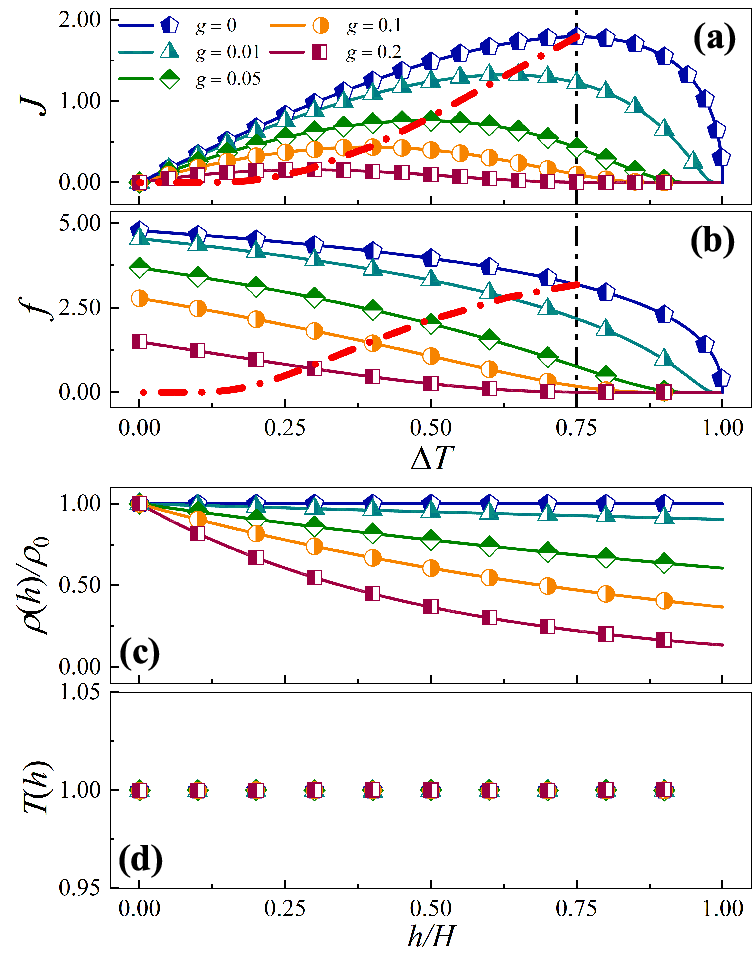}
\caption{
In the integrable case ($\tau = \infty$): (a) heat flux $J$ and (b) collision frequency $f$ versus temperature bias $\Delta T$; (c) normalized number density $\rho(h)/\rho_0$ and (d) local temperature $T(h)$ versus height $h$ at $\Delta T = 0$, for several values of $g$. Here, $\rho_0$ denotes the particle density at zero gravitational potential energy. In panels (a) and (b), solid curves in distinct colors are analytical predictions from Eqs.~(\ref{J11}) and (\ref{f}), respectively; the black dot-dashed line marks the critical bias $(\Delta T)_{\mathrm{cr}} = 0.75$ for $g = 0$, and the red dot-dashed line connects the $(\Delta T)_{\mathrm{cr}}$ values at which NDTR emerges for each $g$. In (c), solid lines show the equilibrium Boltzmann profile $\rho(h) = \rho_0 \exp(-mgh/k_{\mathrm{B}}T)$.
}\label{fig2}
\end{figure}

To enable direct comparison with the $g=0$ result of Ref.~\cite{Luo2022}, we substitute $T_U = T$ and $T_L = T - \Delta T$ into Eq.~(\ref{J11}) and plot $J$ versus $\Delta T$ for several values of $g$. The resulting curves are shown in Fig.~\ref{fig2}(a) as solid lines in distinct colors. The maxima of $J$ at the critical bias $(\Delta T)_{\mathrm{cr}}$---determined by solving $\partial J / \partial(\Delta T) = 0$---are connected by the red dot-dashed line. For $g = 0$, NDTR emerges when $\Delta T > (\Delta T)_{\mathrm{cr}} = 0.75$, as $J$ decreases with increasing $\Delta T$. Crucially, we find that increasing $g$ significantly reduces $(\Delta T)_{\mathrm{cr}}$, thereby greatly expanding the NDTR region. Notably, reversing the temperature gradient (i.e., setting $T_U = T$, $T_L = T + \Delta T$) eliminates NDTR entirely. This occurs because, at fixed $g$, the mean transit time $\langle t_{L\to U} \rangle$ in Eq.~(\ref{J11}) decreases monotonically with $\Delta T$, causing $J$ to increase monotonically.

Next, we elucidate why a larger $g$ reduces $(\Delta T)_{\mathrm{cr}}$. To reach the upper heat bath after reflection from the lower bath, a particle must satisfy $v_y \geq \sqrt{2gH}$ at the lower boundary. As $g$ increases, the required initial upward velocity grows, implying that particles undergo more collisions with the lower bath before reaching the upper bath. Consequently, the mean return time to the upper bath increases. This directly suppresses the collision frequency $f$ of the particles with the upper bath. For clarity, we rewrite the expression for $f$ from Eq.~(6) of Ref.~\cite{Luo2019} as
\begin{equation}\label{f}
  f = \frac{N}{\langle t_{U\to L} \rangle + \langle t_{L\to U} \rangle}.
\end{equation}
This analytical expression is also plotted as the solid curve in Fig.~\ref{fig2}(b), confirming that $f$ decreases with $\Delta T$, and that at fixed $\Delta T$, larger $g$ further diminishes $f$, as expected. Ref.~\cite{Luo2022} established that, for $g=0$, NDTR arises when the inhibitory effect of decreasing $f$ on thermal exchange outweighs the promotional effect of increasing $\Delta T$. Since increasing $g$ amplifies the suppression of $f$, the threshold $(\Delta T)_{\mathrm{cr}}$ at which this dominance occurs shifts to smaller values. Hence, gravity actively promotes NDTR by lowering $(\Delta T)_{\mathrm{cr}}$.

\begin{figure*}
\centering
\includegraphics[width=17cm]{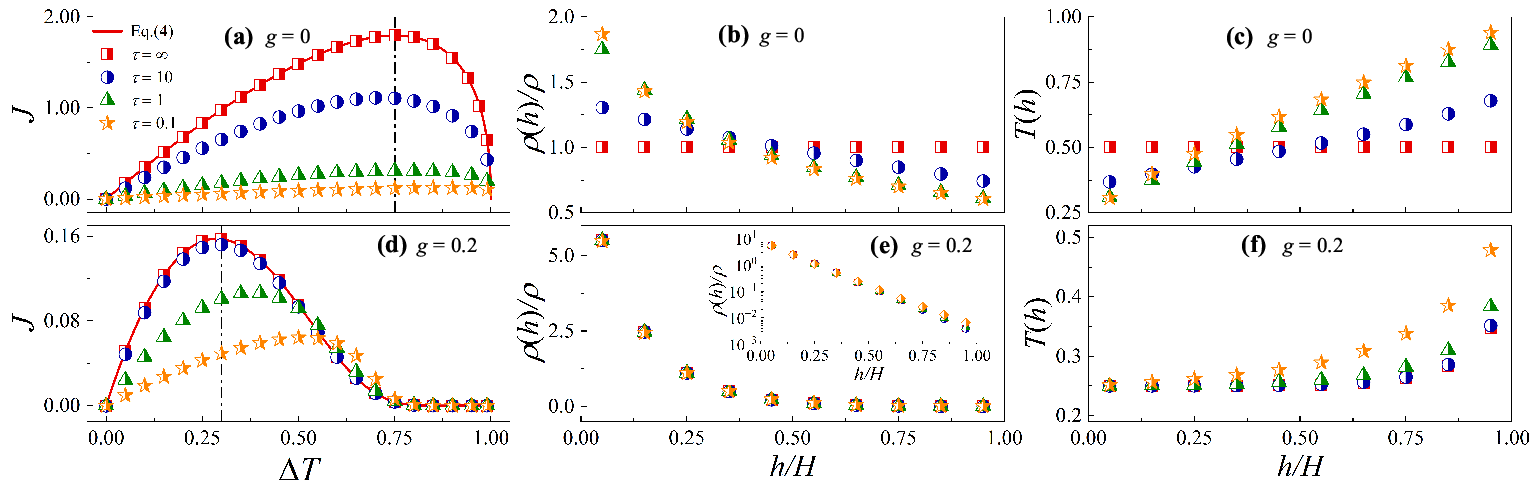}
\caption{
Numerical results for varying MPC interaction time interval $\tau$:
(a, d) heat flux $J$ versus temperature bias $\Delta T$ at $g = 0$ and $g = 0.2$, respectively;
(b, e) particle number density $\rho(h)$ versus height $h$ at $\Delta T = 0.75$, for $g = 0$ and $g = 0.2$;
(c, f) local temperature $T(h)$ versus height $h$ under identical conditions with (b, e).
In (a) and (d), the red curve is the analytical prediction from Eq.~(\ref{J11}); the black dot-dashed line marks the critical bias $(\Delta T)_{\mathrm{cr}}$ in the noninteracting limit ($\tau = \infty$).
The semi-log inset in (e) shows exponential decay of $\rho(h)$ with height, consistent with the Boltzmann distribution.
}
\label{fig3}
\end{figure*}


\textit{Numerical results.} To validate the analytical results and illustrate their physical relevance, we first consider the noninteracting (integrable) case. In Figs.~\ref{fig2}(a) and \ref{fig2}(b), the theoretical curves for the heat flux [Eq.~(\ref{J11})] and collision frequency [Eq.~(\ref{f})] are compared with numerical simulation data (symbols). Excellent agreement is found across the entire parameter range, corroborating the validity of our analytical framework. Furthermore, to elucidate the role of gravity in equilibrium, Figs.~\ref{fig2}(c) and \ref{fig2}(d) display the height-dependent particle number density $\rho(h)$ and local temperature $T(h)$ at $\Delta T = 0$, for several values of $g$. As expected, $\rho(h)$ obeys the Boltzmann distribution $\rho(h) = \rho_0 \exp\!\left(-\frac{mgh}{k_{\mathrm{B}} T}\right)$. Meanwhile, $T(h) = T = 1$ is uniform throughout the system, confirming the expected homogeneous isothermal state in thermal equilibrium.\par

We now turn to interacting systems with collisions to investigate the dependence of the NDTR mechanism on interaction strength. Here, the time interval $\tau$ between successive collisions serves as a control parameter for tuning the interaction strength. In the noninteracting limit, $\tau = \infty$; thus, for interacting systems, a smaller $\tau$ corresponds to stronger interactions. As shown in Fig.~\ref{fig3}(a), for $g = 0$, decreasing $\tau$ reduces the size of the NDTR region, which vanishes entirely at sufficiently small $\tau$. This indicates that the heat-bath-induced NDTR mechanism breaks down under relatively strong interactions (e.g., $\tau = 0.1$). However, as seen in Fig.~\ref{fig3}(d), for $g = 0.2$, although decreasing $\tau$ increases the critical temperature difference $(\Delta T)_{\mathrm{cr}}$, NDTR persists even at $\tau = 0.1$. This result demonstrates that, in the presence of gravity, the mechanism underlying NDTR remains effective in systems with relatively strong interactions.\par

Next, we explain why this mechanism breaks down in the absence of gravity under strong interactions, yet remains effective when gravity is present. As shown in Figs.~\ref{fig3}(b) and \ref{fig3}(c), for $g = 0$, decreasing $\tau$ increases $\rho(h)$ at the lower (cold bath) boundary and drives $T(h)$ toward a linear profile characteristic of Fourier's law. This indicates that, for $\tau = 0.1$, the contribution of increasing $\Delta T$ to the heat flux becomes dominant, restoring the conventional monotonic increase of $J$ with $\Delta T$. As explained in~\cite{Luo2022}, this behavior arises because stronger interactions enhance momentum exchange between particles, causing low-velocity particles reflected from the cold bath to gain speed and thereby reach the hot bath more quickly, thus enhancing inter-bath heat transfer. In contrast, Fig.~\ref{fig3}(e) shows that, for $g = 0.2$, $\rho(h)$ decreases exponentially with increasing $h$, and curves for different $\tau$ values nearly overlap; correspondingly, the $T(h)$ curves in Fig.~\ref{fig3}(f) not only exhibit reduced magnitudes but also display a nonlinear profile, markedly differing from the $g = 0$ case. This occurs because, in the presence of gravity, although strong interactions further increase the velocities of particles at the cold end, the kinetic energy gained by most of these particles is still insufficient to overcome the gravitational potential barrier and reach the hot end of the system, thereby preserving the NDTR effect even under strong interactions.\par

\begin{figure}
\centering
\includegraphics[width=7.5cm]{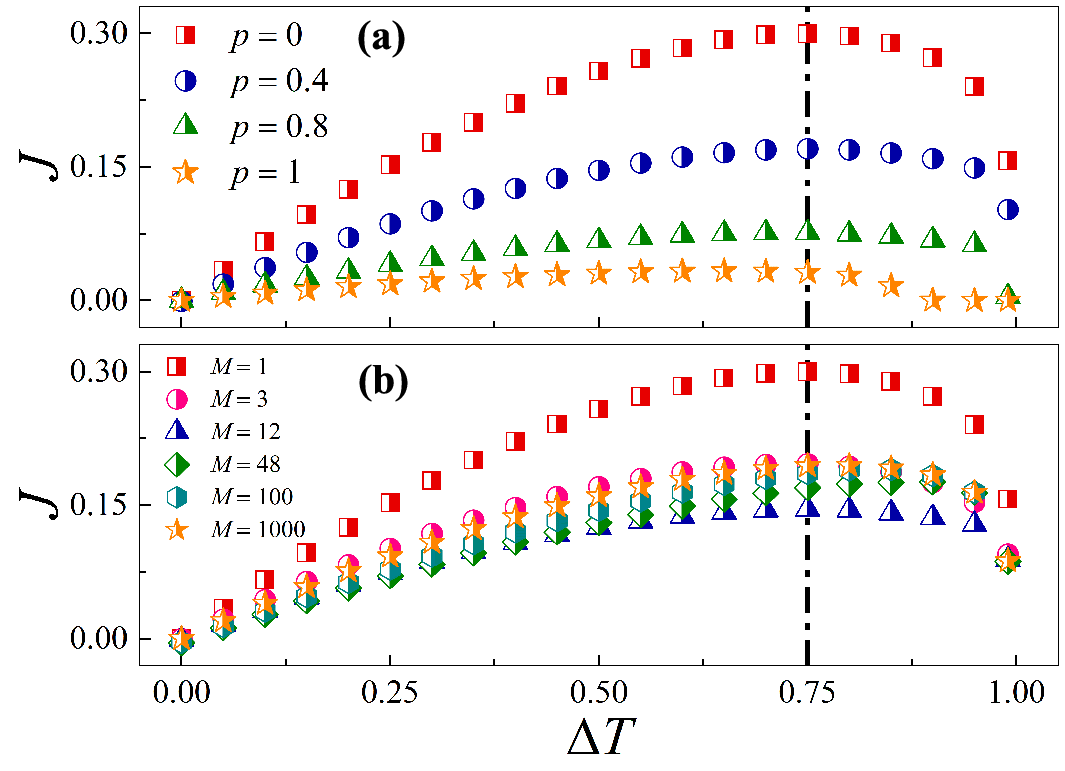}
\caption{Heat flux $J$ versus temperature bias $\Delta T$ for binary fluid systems: (a) varying $p$ at fixed $M=12$; (b) varying $M$ at fixed $p=0.5$. The dot-dashed line in (a) and (b) indicates the critical temperature bias $(\Delta T)_{\mathrm{cr}}$ for the onset of NDTR. Parameters: $g=0.01$, $\tau=1$.}
\label{fig4}
\end{figure}

To further demonstrate the universality of this mechanism under gravity, we apply the simulation method described in~\cite{Gompper2009}, confirming its applicability to binary mixtures. We consider a simple binary fluid composed of particles with two distinct masses, $m$ and $M$, where the heavier mass $M$ occurs with probability $p$. For simplicity, we set $m=1$, so that $M$ represents the mass ratio $M/m$. Results for varying $p$ at fixed $M=12$ and varying $M$ at fixed $p=0.5$, with $g=0.01$ and $\tau=1$, are presented in Figs.~\ref{fig4}(a) and~\ref{fig4}(b), respectively. Our data show that the mechanism under gravity operates both in pure fluids ($p=0$ or $1$, or $M=1$) and in mixed binary fluids across a range of $p$ and $M$. Crucially, for fixed interaction strength ($\tau=1$), the NDTR region remains unchanged upon increasing $p$ and $M$, indicating robustness across diverse binary mixtures. These results provide insight into, and may enable control of, heat transport in binary fluid mixtures under gravity~\cite{Thorsten2001,van1997,Squires2005,Scholz2017}.



\textit{Discussions and conclusions.} In summary, we have investigated the role of gravity in the emergence and characteristics of NDTR in a fluid model simulated using MPC dynamics. We derive an analytical expression for the heat flux that agrees well with simulations, demonstrating that the temperature bias required for NDTR is significantly reduced when gravity is aligned with the thermodynamic force. Furthermore, gravity extends the applicability of the heat-bath-induced NDTR mechanism to systems with stronger interactions, beyond the weakly interacting regime in which it was originally established. Notably, this mechanism under gravity remains robust across a variety of fluid mixtures.\par

The case of gravity opposing the thermodynamic force is studied in Sec.~II of the Supplemental Material. In this configuration, while the mechanism also works in fluid mixtures, the threshold temperature bias for NDTR is substantially increased, and gravity no longer extends the mechanism to stronger interactions. These findings highlight the directional sensitivity of gravitational control and underscore the generality and tunability of NDTR in nonequilibrium fluids.\par

The influence of gravity on heat transport is central to both fundamental nonequilibrium physics and potential thermal management applications. The MPC dynamics with gravity presented here provides a computationally efficient model for exploring such effects. Extensions to higher-dimensional systems could enable the study of Rayleigh-B\'{e}nard flow~\cite{Stefanov2022,2022Stefanov}, thermal rectification~\cite{2024zou}, and dimension-dependent heat conduction~\cite{Lepri2016}. Furthermore, incorporating thermochemical baths---allowing both heat and particles exchange with the environment~\cite{Benenti2014,Luo2020}---would enable the study of coupled heat and particle transport, offering insights into thermoelectric-like effects and potential designs for fluidic thermal devices.\par

Fluidic TCDs---including thermal switches, diodes, and regulators---have been implemented across diverse length and temperature scales~\cite{Klinar2021}. However, unlike electronic systems, fluidic thermal engineering lacks a functional analogue to the transistor. The NDTR demonstrated here provides a mechanism for strong nonlinearity and signal gain, essential for realizing a fluidic thermal transistor. This paves the way for designing such a transistor and other more complex fluidic TCDs, as previously achieved in lattice systems and quantum systems~\cite{Li2012,Landi2022}. The present work will stimulate further theoretical and experimental investigation of NDTR in fluidic systems.

\textit{Acknowledgment.} We are grateful to an anonymous referee for their helpful comments and suggestions. We acknowledge support by the National Natural Science Foundation of China (Grants No.12475034 and No.12105049) and the Natural Science Foundation of Fujian Province (Grant No.2023J05100).

\textit{Data availability.} The data that support the findings of this article are not publicly available. The data are available
from the authors upon reasonable request.

\bibliography{paper}

\appendix

\begin{widetext}

\clearpage
\section*{Supplemental Material for ``Gravity-Induced Modulation of Negative Differential Thermal Resistance in Fluids"}

\quad Here we present the detailed derivation of the heat flux and collision frequency for the fluid models under study in the presence of gravity. For clarity, this part focuses exclusively on the analytical derivation of Eqs.~(4)-(6) in the main text.

In addition, we consider a complementary configuration in which gravity acts opposite to the thermodynamic force, and investigate the corresponding negative differential thermal resistance (NDTR), comparing the results with those reported in the main text.
\section{Derivation of heat flux and collision frequency}\label{sec:AppA}

\subsection{Derivation of heat flux}
The fluid model shown in Fig.~\ref{sfig1} corresponds to an integrable case in which interparticle interactions are neglected. In this case, the steady-state heat flux flowing across the system from the lower to the upper boundary is given by
\begin{equation}\label{J}
J=\frac{N}{L}\left|\frac{\langle E_{U\rightarrow L}\rangle-\left(\langle E_{L\rightarrow U}\rangle-mgH\right)}{\langle t_{U\rightarrow L}\rangle+\langle t_{L\rightarrow U}\rangle}\right|,
\end{equation}
where $N$ is the number of particles in the system, $L$ is the system length, $mgH$ represents the work done by a particle against gravity when moving upward, and $\langle E_{U\rightarrow L}\rangle$, $\langle t_{U\rightarrow L}\rangle$ denote, respectively, the average energy transferred and the average transit time for a particle traveling from the upper to the lower boundary. Similarly, $\langle E_{L\rightarrow U}\rangle$ and $\langle t_{L\rightarrow U}\rangle$ refer to the corresponding quantities for the reverse journey (from lower to upper). Note that a particle launched from the upper boundary reaches the lower boundary directly, whereas a particle launched from the lower boundary may undergo one or more collisions with the lower boundary before reaching the upper boundary, due to the influence of gravity.\par

In the following, we first compute $\langle E_{U\rightarrow L}\rangle$ and $\langle E_{L\rightarrow U}\rangle$.
The average energy transferred by a particle during a journey from the upper to the lower boundary is given by
\begin{align}\label{E_UD}
\langle E_{U\rightarrow L}\rangle
= \int_{0}^{\infty} \frac{1}{2} m v_y^{2} f_U(v_y) \, dv_y
  + \int_{-\infty}^{\infty} \frac{1}{2} m v_x^{2} f_U(v_x) \, dv_x
= \frac{3}{2} k_{\mathrm{B}} T_U .
\end{align}
For a particle to travel from the lower to the upper boundary, its $y$-component of velocity must satisfy $v_y \geq \sqrt{2gH}$. Hence, the average energy transferred during such a journey is
\begin{align}\label{E_DU}
\langle E_{L\rightarrow U}\rangle
= \frac{\int_{\sqrt{2gH}}^{\infty} \frac{1}{2} m v_y^{2} f_L(v_y) \, dv_y}
       {\int_{\sqrt{2gH}}^{\infty} f_L(v_y) \, dv_y}
  + \int_{-\infty}^{\infty} \frac{1}{2} m v_x^{2} f_L(v_x) \, dv_x
= \frac{3}{2} k_{\mathrm{B}} T_L + mgH .
\end{align}

Next, we compute $\langle t_{U\rightarrow L}\rangle$, the average time a particle takes to travel from the upper to the lower boundary. The transit time for a particle moving from the upper to the lower boundary is given by
\begin{equation}\label{Eqtud}
t_{U\rightarrow L} = \frac{\sqrt{v_{y,U}^{2} + 2gH} - v_{y,U}}{g},
\end{equation}
which follows from the energy conservation relation
\begin{equation}\label{Eq2}
\frac{1}{2} m v_{y,L}^{2} = \frac{1}{2} m v_{y,U}^{2} + mgH,
\end{equation}
and the dynamic equation
\begin{equation}\label{Eq3}
v_{y,L} = v_{y,U} + g t_{U\rightarrow L},
\end{equation}
where $v_{y,U}$ and $v_{y,L}$ denote the $y$-components of the particle's velocity at the upper and lower boundaries, respectively.
The ensemble-averaged transit time is then
\begin{align}\label{t_UD}
\langle t_{U\rightarrow L}\rangle
&= \int_{0}^{\infty} t_{U\rightarrow L} \, f_U(v_y) \, dv_y \notag \\
&= \frac{1}{g} \sqrt{\frac{\pi k_{\mathrm{B}} T_U}{2m}} \left\{
e^{\frac{mgH}{k_{\mathrm{B}} T_U}} \left[ 1 - \mathrm{Erf}\!\left( \sqrt{\frac{mgH}{k_{\mathrm{B}} T_U}} \right) \right] - 1
\right\} + \sqrt{\frac{2H}{g}} .
\end{align}

Then, we compute $\langle t_{L\rightarrow U}\rangle$.
The average transit time for journeys from the lower to the upper boundary is not straightforward, as a particle may undergo multiple collisions with the lower boundary before reaching the upper one.
Let $p_n$ denote the probability that a particle launched from the lower boundary experiences exactly $n-1$ collisions with the lower boundary prior to reaching the upper boundary, and let $t_n$ be the corresponding travel time for such a journey. The ensemble-averaged transit time is then given by
\begin{equation}\label{Eqtdu}
\langle t_{L\rightarrow U}\rangle = \sum_{n=1}^{\infty} p_n t_n,
\end{equation}
where the sum runs over all admissible journeys. For a particle starting at the lower boundary, let $p_{L\rightarrow U}$ and $p_{L\rightarrow L}$ denote the probabilities of next striking the upper or lower boundary, respectively, and let $t_{L\rightarrow U}$ and $t_{L\rightarrow L}$ be the associated transit times. Then,
\begin{equation}\label{Eqpn}
p_n = p_{L\rightarrow L}^{\,n-1} \, p_{L\rightarrow U},
\end{equation}
\begin{equation}\label{Eqtn}
t_n = (n-1)\, t_{L\rightarrow L} + t_{L\rightarrow U}.
\end{equation}

From the definitions above, we obtain
\begin{equation}\label{Eqpud}
p_{L\rightarrow L} = \int_{0}^{\sqrt{2gH}} f_L(v_y) \, dv_y = 1 - e^{-\frac{mgH}{k_{\mathrm{B}} T_L}},
\end{equation}
\begin{equation}\label{Eqpdu}
p_{L\rightarrow U} = \int_{\sqrt{2gH}}^{\infty} f_L(v_y) \, dv_y = e^{-\frac{mgH}{k_{\mathrm{B}}T_L}},
\end{equation}
\begin{align}\label{EqtDD}
t_{L\rightarrow L}
&= \frac{\int_{0}^{\sqrt{2gH}} t_{L\rightarrow L} \, f_L(v_y) \, dv_y}{p_{L\rightarrow L}} \notag \\
&= \frac{-2\sqrt{2gH}\, e^{-\frac{mgH}{k_{\mathrm{B}} T_L}} + \sqrt{\frac{2\pi k_B T_L}{m}} \, \mathrm{Erf}\!\left(\sqrt{\frac{mgH}{k_{\mathrm{B}} T_L}}\right)}{g\left(1 - e^{-\frac{mgH}{k_{\mathrm{B}} T_L}}\right)},
\end{align}
\begin{align}\label{EqtDU}
t_{L\rightarrow U}
&= \frac{\int_{\sqrt{2gH}}^{\infty} t_{L\rightarrow U} \, f_L(v_y) \, dv_y}{p_{L\rightarrow U}} \notag \\
&= \sqrt{\frac{2H}{g}} - \frac{1}{g} \sqrt{\frac{\pi k_{\mathrm{B}} T_L}{2m}} \left[ 1 - e^{\frac{mgH}{k_{\mathrm{B}} T_L}} \, \mathrm{Erfc}\!\left( \sqrt{\frac{mgH}{k_{\mathrm{B}} T_L}} \right) \right].
\end{align}

In Eqs.~(\ref{EqtDD}) and (\ref{EqtDU}), the expression $t_{L\rightarrow L} = \frac{2v_{y,L}}{g}$ follows from the free-fall motion of a particle, while $t_{L\rightarrow U} = \frac{v_{y,L} - \sqrt{v_{y,L}^2 - 2gH}}{g}$ is derived analogously to Eq.~(\ref{Eqtud}).
Substituting Eqs.~(\ref{Eqpn})--(\ref{EqtDU}) into Eq.~(\ref{Eqtdu}) yields
\begin{align}\label{EqtDU1}
\langle t_{L\rightarrow U}\rangle
&= \sum_{n=1}^{\infty} p_n t_n
= \frac{p_{L\rightarrow L}}{p_{L\rightarrow U}} \, t_{L\rightarrow L} + t_{L\rightarrow U} \notag \\
&= \frac{1}{g} \sqrt{\frac{\pi k_{\mathrm{B}} T_L}{2m}} \left\{
e^{\frac{mgH}{k_{\mathrm{B}} T_L}} \left[ \mathrm{Erf}\!\left( \sqrt{\frac{mgH}{k_{\mathrm{B}} T_L}} \right) + 1 \right] - 1
\right\} - \sqrt{\frac{2H}{g}} .
\end{align}
Equation~(\ref{EqtDU1}) has a clear physical interpretation: for a particle launched from the lower boundary and eventually reaching the upper boundary, the mean number of collisions with the lower boundary is $p_{L\rightarrow L}/p_{L\rightarrow U}$. Consequently, the average transit time naturally separates into two contributions, as reflected on the right-hand side of Eq.~(\ref{EqtDU1}).

The above derivations can be readily generalized to one- or three-dimensional systems.
Summarizing these results, the heat flux expression, Eq.~(\ref{J}), generalized to $d$ spatial dimensions, takes the form
\begin{equation}\label{J1}
J = \rho H \, \frac{1+d}{2} \left| \frac{k_{\mathrm{B}} (T_U - T_L)}{\langle t_{U\rightarrow L} \rangle + \langle t_{L\rightarrow U} \rangle} \right|,
\end{equation}
where $d \in \{1,2,3\}$ denotes the spatial dimension, and $\rho = N/(LH)$ is the mean particle number density.\par

\subsection{Derivation of collision frequency}
The collision frequency for a single particle striking the upper boundary is defined as $f^{(1)}=1/(\langle t_{L\rightarrow U}\rangle+\langle t_{U\rightarrow L}\rangle)$. For a system of $N$ non-interacting particles, the total collision frequency with the upper boundary is then
\begin{equation}\label{Eqf}
f = N f^{(1)} = \frac{N}{\langle t_{L\rightarrow U} \rangle + \langle t_{U\rightarrow L} \rangle}.
\end{equation}

\section{NDTR of fluids with gravity opposing the thermodynamic force}\label{sec:AppB}

\begin{figure}[H]
\centering
\includegraphics[width=3.2cm]{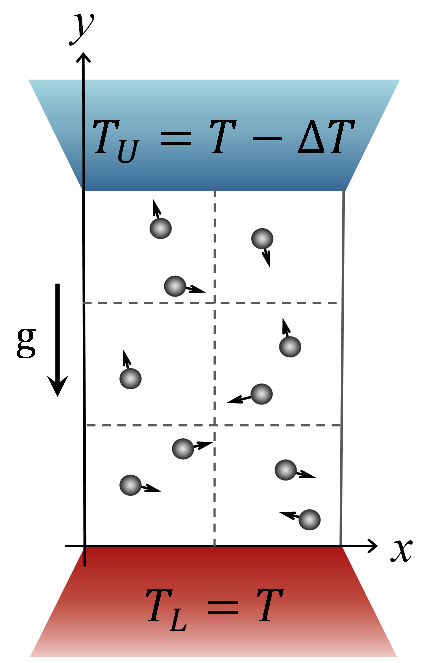}
\caption{Schematic of the 2D multiparticle collision fluid model with gravity counter to the thermodynamic force.}
\label{sfig1}
\end{figure}

In this section, we examine the heat-bath-induced NDTR mechanism in MPC fluids under a configuration where gravity opposes the thermodynamic force.
Specifically, we set $T_U = T - \Delta T$ and $T_D = T$, as illustrated in Fig.~\ref{sfig1}, with all other simulation parameters held fixed at their values from the main text.
This enables a direct and physically meaningful comparison between the two configurations.\par

\begin{figure}
\centering
\includegraphics[width=14cm]{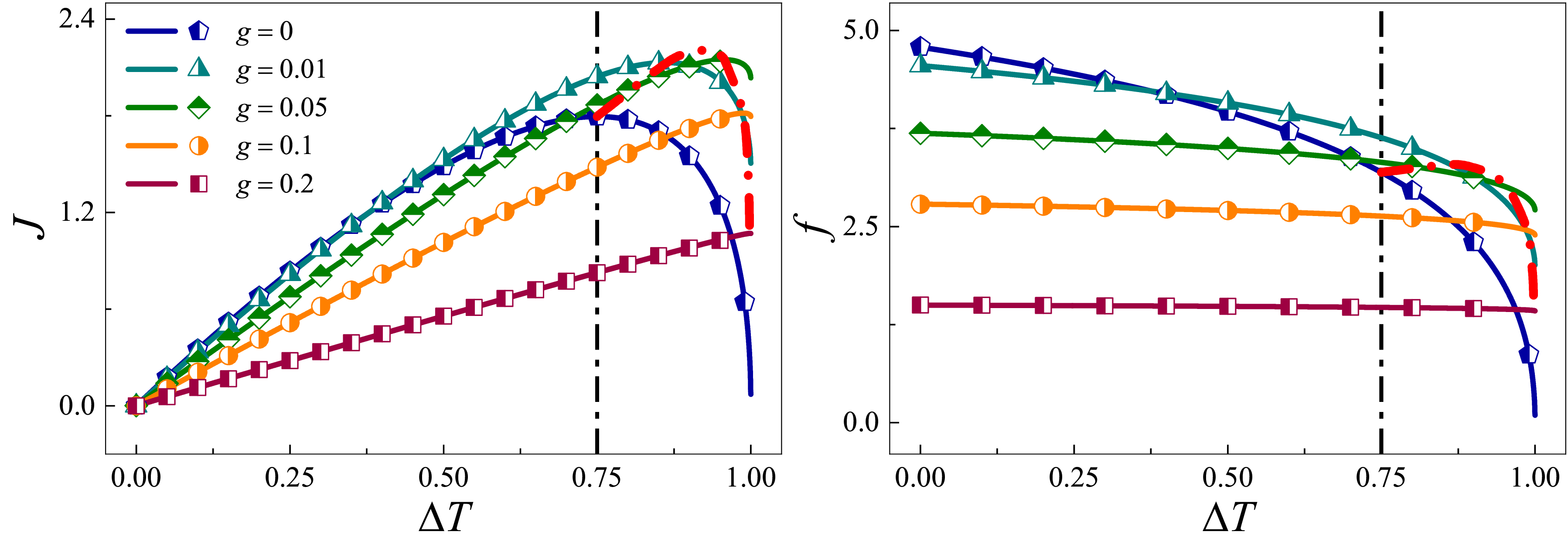}
\caption{In the integrable case, (a) heat flux $J$ and (b) collision frequency $f$ versus temperature bias $\Delta T$, for various $g$. Solid colored curves in (a) and (b) show the analytical results from Eqs.~(\ref{J1}) and (\ref{Eqf}), respectively. For reference, the black dot-dashed line marks $(\Delta T)_{\mathrm{cr}} = 0.75$ at $g = 0$; the red dot-dashed line traces the $g$-dependence of $(\Delta T)_{\mathrm{cr}}$, indicating the onset of NDTR. }
\label{sfig2}
\end{figure}

In the integrable case, Fig.~\ref{sfig2}(a) shows the heat flux $J$ versus temperature bias $\Delta T$ for various $g$, with solid colored curves representing the analytical results from Eq.~(\ref{J1}). When gravity opposes the thermodynamic force, increasing $g$ raises $(\Delta T)_{\mathrm{cr}}$ at which NDTR emerges, thereby narrowing the NDTR region; for $g > 0.1$, this region vanishes entirely. To elucidate this behavior, Fig.~\ref{sfig2}(b) plots the collision frequency $f$ of particles with the upper heat bath. For $g = 0.01$ and $g = 0.05$, $f$ decreases with $\Delta T$ and exhibits a sharp increase when $\Delta T> 0.75$ relative to the $g = 0$ case. This enhancement in $f$ raises $(\Delta T)_{\mathrm{cr}}$, suppressing NDTR onset. Moreover, for $g > 0.1$, $f$ becomes independent of $\Delta T$; consequently, $J$ is no longer influenced by $f$ and depends solely on $\Delta T$. As a result, $J$ increases monotonically with $\Delta T$, and thus NDTR no longer occurs, as shown in Fig.~\ref{sfig2}(a).\par

Figures~\ref{sfig3}(a) and~\ref{sfig3}(d) illustrate the dependence of the heat-bath-induced NDTR mechanism on interaction strength in the absence and presence of gravity, respectively. Consistent with the $g = 0$ case, decreasing the collision time step $\tau$ shrinks the NDTR region at $g = 0.01$; this region vanishes entirely for sufficiently small $\tau$ (see $\tau = 0.1$ as reference). This indicates that, under weak gravity, the mechanism also breaks down for systems with strong interactions, in contrast to the behavior reported in the main text. To elucidate this transition, we plot the particle number density $\rho(h)$ and temperature profile $T(h)$ for different $\tau$. As shown in Figs.~\ref{sfig3}(e) and~\ref{sfig3}(f), for $g = 0.01$, decreasing $\tau$ increases $\rho(h)$ near the upper (cold) bath and drives $T(h)$ toward a linear profile characteristic of Fourier's law---consistent with the $g = 0$ case (Figs.~\ref{sfig3}(b) and~\ref{sfig3}(c)). Thus, weak gravity does not modify the mechanism by which strong interactions suppress NDTR; the disappearance of NDTR with decreasing $\tau$ is entirely caused by the increased interaction strength, which enhances the velocities of low-velocity particles reflected from the cold bath and thereby facilitates heat transfer across the baths, as explained in the main text.

\begin{figure}
\centering
\includegraphics[width=16cm]{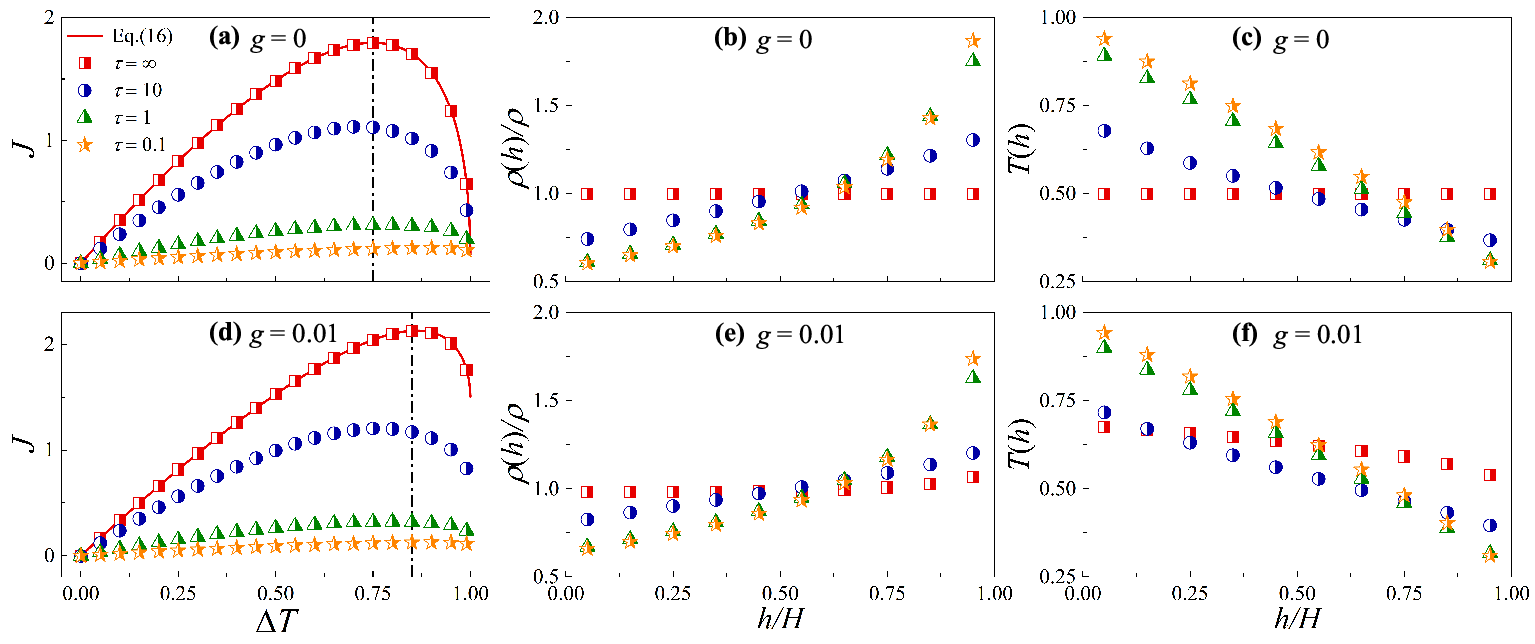}
\caption{Results are shown for varying collision time interval $\tau$: heat flux $J$ versus $\Delta T$ at (a) $g = 0$ and (d) $g = 0.01$; number density $\rho(h)$ versus $h$ at $\Delta T = 0.75$ for (b) $g = 0$ and (e) $g = 0.01$; local temperature $T(h)$ versus $h$ at $\Delta T = 0.75$ for (c) $g = 0$ and (f) $g = 0.01$. In (a) and (d), the red curve shows Eq.~(\ref{J1}), and the dot-dashed line indicates $(\Delta T)_{\mathrm{cr}}$ for $\tau = \infty$.}
\label{sfig3}
\end{figure}

To further demonstrate the applicability of the NDTR mechanism to binary mixtures in this configuration, we perform an analysis analogous to that in the main text on a simple binary fluid. Results for varying the heavy-particle probability $p$ at fixed mass ratio $M = 12$, and varying $M$ at fixed $p = 0.5$, with $g = 0.01$ and $\tau = 10$, are presented in Figs.~\ref{sfig4}(a) and~\ref{sfig4}(b), respectively. The data show that the mechanism under gravity operates in binary mixtures across a range of $p$ and $M$. However, for fixed interaction strength ($\tau = 10$) and $M = 12$, the NDTR region in Fig.~\ref{sfig4}(a) gradually diminishes and eventually vanishes as $p$ increases from 0 to 1, indicating that its robustness is limited when heavy particles dominate the mixture.

\begin{figure}
\centering
\includegraphics[width=14cm]{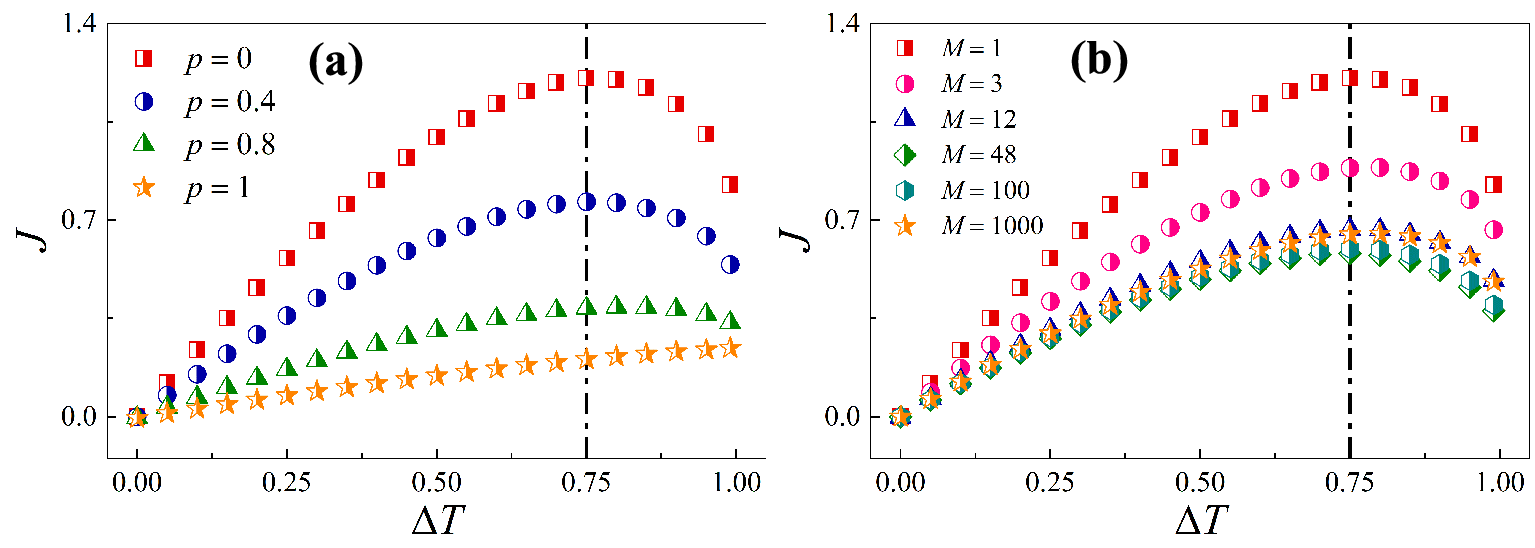}
\caption{$J$ as a function of $\Delta T$ for binary fluid systems: (a) varying $p$ at fixed $M=12$; (b) varying $M$ at fixed $p=0.5$. For reference, the dot-dashed line in (a) and (b) indicates $(\Delta T)_{\mathrm{cr}}$ for the onset of NDTR. Parameters: $g=0.01$, $\tau=10$.}
\label{sfig4}
\end{figure}
\end{widetext}
\end{document}